\documentclass[10pt,letterpaper,aps,pra,twocolumn,superscriptaddress]{revtex4-1}
\usepackage[latin1]{inputenc}
\usepackage{amsmath}
\usepackage{amsfonts}
\usepackage{amssymb}
\usepackage{bm}
\usepackage[pdftex]{graphicx}
\usepackage[colorinlistoftodos]{todonotes}

\newcommand{\mean}[1]{\left\langle #1 \right\rangle}

\newcommand{\ra}{\rangle}
\newcommand{\be}{\begin{equation}}
\newcommand{\ee}{\end{equation}}

\begin{document}
\title{Rapid estimation of drifting parameters in continuously measured quantum systems}
\author{Luis Cortez$^*$}
\affiliation{Department of Physics and Astronomy, University of Rochester, Rochester, New York 14627, USA}
\affiliation{Center for Coherence and Quantum Optics, University of Rochester, Rochester, New York 14627, USA}
\affiliation{Facultad de Ciencias Fisico Matematicas, Universidad Autonoma de Nuevo Leon, Nuevo Leon 66455, Mexico}

\author{Areeya Chantasri$^*$}
\affiliation{Department of Physics and Astronomy, University of Rochester, Rochester, New York 14627, USA}
\affiliation{Center for Coherence and Quantum Optics, University of Rochester, Rochester, New York 14627, USA}

\author{Luis Pedro Garc\'ia-Pintos}
\affiliation{Institute for Quantum Studies, Chapman University, Orange, California 92866, USA}

\author{Justin Dressel}
\affiliation{Institute for Quantum Studies, Chapman University, Orange, California 92866, USA}
\affiliation{Schmid College of Science and Technology, Chapman University, Orange, California 92866, USA}

\author{Andrew N. Jordan}
\affiliation{Department of Physics and Astronomy, University of Rochester, Rochester, New York 14627, USA}
\affiliation{Center for Coherence and Quantum Optics, University of Rochester, Rochester, New York 14627, USA}
\affiliation{Institute for Quantum Studies, Chapman University, Orange, California 92866, USA}
\date{\today}

\begin{abstract}
We investigate the determination of a Hamiltonian parameter in a quantum system undergoing continuous measurement. We demonstrate a computationally rapid yet statistically optimal method to estimate an unknown and possibly time-dependent parameter, where we maximize the likelihood of the observed stochastic readout. By dealing directly with the raw measurement record rather than the quantum state trajectories, the estimation can be performed while the data is being acquired, permitting continuous tracking of the parameter during slow drifts in real time. Furthermore, we incorporate realistic nonidealities, such as decoherence processes and measurement inefficiency. As an example, we focus on estimating the value of the Rabi frequency of a continuously measured qubit, and compare maximum likelihood estimation to a simpler fast Fourier transform. Using this example, we discuss how the quality of the estimation depends on both the strength and duration of the measurement; we also discuss the trade-off between the accuracy of the estimate and the sensitivity to drift as the estimation duration is varied.
\end{abstract}

\maketitle

\section{Introduction}
The problem of accurately measuring unknown parameters in an experimental system is of both fundamental and practical importance. The process of determining the parameters of the experiment serves as a valuable calibration of the experiment, and also determines the limitations of experimental accuracy. The understanding of how to minimize parameter uncertainties given a variety of possible measurement strategies has been developed into the science of quantum metrology over the past several decades \cite{Giovannetti2004,Giovannetti2011}.

The usual approach taken in the laboratory is to repeatedly perform the following sequence of operations: prepare a quantum state, let it evolve unitarily in a way that depends on the unknown parameter, and then perform a (potentially unsharp) measurement to extract information about the parameter of interest. The concept of the quantum Fisher information \cite{wisebook} is fundamental to this approach, since it sets the bound on the minimum variance for all possible estimation strategies based on this final measurement, the so-called \emph{Cram\'er-Rao bound}. However, there are other methods for estimating such an unknown parameter that go beyond the prepare-evolve-measure paradigm, which may be beneficial under certain circumstances. One such scenario is when the parameter changes slowly in time such that this variation cannot be predicted in advance, as is common in experimental laboratories (e.g., from thermal fluctuations). In this case, it is beneficial to be able to \emph{continuously track} the changing parameter as it evolves in time, to sense when and how it is changing. Given knowledge of how the parameters are changing, introducing feedback control to stabilize the parameter then becomes possible \cite{Vijay2012}. Such a situation brings into play the physics of open quantum systems and how they relate to metrology \cite{Escher2011,Demkowicz-Dobrzanski2012,Tsang2013,Kolodynski2013,Alipour2014}.  

To continuously track the changes of a parameter in time, it is natural to consider measurements that are also continuous in time. In order to increase the speed of the estimation, the technique builds upon prior parameter information obtained from an initial broad system characterization. Assuming relatively slow drift of the parameter, the initial characterization narrows the search region of subsequent repeated estimations in real-time as a single noisy measurement record is monitored.

Previously, the physics of parameter estimation using continuous measurements has been analyzed by Ralph, Jacobs, and Hill \cite{ralph2011frequency}.  They numerically integrated a stochastic master equation to estimate the unknown frequency of a Hamiltonian drive for a qubit. Our work is closely related to theirs, and builds off of it. Klaus M{\o}lmer and collaborators have also developed parameter estimation methods using continuous quantum measurements. These relate to how the parameter estimation can be carried out by Bayesian estimation in solving stochastic master equations \cite{KM1,KM2}, and how Fisher information is degraded in the quantum Zeno regime \cite{KM3}. The potential use of continuous measurements for quantum state tomography is also starting to be explored \cite{smith2006efficient,smith2013quantum,six2016quantum}.

In this work, we consider a quantum system undergoing such a continuous measurement, which gives rise to a stochastic measurement record that can be monitored in time.  We wish to analyze this data to extract the value of an unknown parameter in a way that is both computationally rapid and statistically efficient to permit estimation on a short enough time scale where feedback control becomes possible to correct drift. For specificity, we focus on the determination of an unknown and drifting Rabi drive for the qubit. There are a number of open problems in this area which we now consider: How can one minimize the complexity of maximum likelihood algorithms so they are computationally fast?  Is it possible to work only with raw measurement data, so numerical implementations of quantum filters that estimate the quantum state dynamically are not needed?  Is it possible to incorporate numerically more efficient methods to narrow down the parameter search space? Can existing methods be generalized to account for experimental nonidealities, such as additional dephasing, detector inefficiency and energy relaxation? In this paper we work toward solutions to these problems, and outline how they can be experimentally implemented in superconducting circuits, among other possibilities.  Continuous measurement with superconducting circuits have a proven ability to accurately track the quantum state in time \cite{murc13traj} with excellent agreement with predicted statistics \cite{webe14}.

Our basic insight is to speed up the estimation protocol by avoiding the numerical integration of the stochastic master equation.  Rather, we construct effective propagators directly from the observed measurement record that can be used in the maximum likelihood algorithm.
These effective propagators use measurement operators with the sequence of digitized measurement results from, e.g., a homodyne measurement, together with unitary matrices with unknown Rabi drive frequency.  These can all be evaluated numerically using the particular realization of the stochastic measurement results.
By simple multiplication of the composite matrices in the effective propagator, a suitable likelihood function is straightforwardly constructed with the initial state, which can then be maximized. Such maximum likelihood estimation (MLE) saturates the Cram\'er-Rao bound. Combining this MLE technique with an initial fast Fourier transform (FFT) technique, (which identifies a range of prior Rabi frequencies) provides a significant speed up by decreasing the number of trial frequencies needed for MLE. We further generalize this method to incorporate realistic non-idealities to prepare this method for experimental implementation in superconducting circuit architectures, where continuous homodyne and heterodyne measurement are now routinely carried out. We believe this method will be suitable to be directly programmed into a field-programmable gate array (FPGA) for rapid near-real-time implementation. Finally, we illustrate how this method can be applied to a time-varying, unknown Rabi drive, and show that we can accurately track even irregular motion in time.  Notably, a \emph{projective} measurement version of this task has been accomplished by Shulman, Harvey, Nichol, {\it et al.} using such an FPGA in a triplet/singlet spin qubit in order to detect how the surrounding nuclear magnetic field was changing, and incorporate feedback to prolong the qubit dephasing time \cite{shulman2014suppressing}. We thus present our own analysis of the projective equivalent with fixed spacing between measurements in the Appendix as a comparison to the present work.

The paper is organized as follows. In Sec.~II, we outline our strategy for determining the value of a static Hamiltonian parameter by maximizing the likelihood of observing a particular stochastic measurement record. We focus on the example of a driven qubit, where the parameter to be estimated is an initially static Rabi frequency. We then compare the maximum likelihood approach to a simpler FFT, which can be used to help identify a suitable frequency range for subsequent maximum likelihood estimation, and comment on the relative computational efficiency of each method. 
In Sec. III we generalize the static estimation method to a dynamic estimation method that is able to track arbitrary time-dependent parameters. Setting a desired estimation precision then specifies the time resolution for the tracking of drift. We demonstrate that we are able to accurately track dynamical parameters using this method. We conclude in Sec. IV. We also provide an Appendix that includes an analytic treatment of the maximum likelihood frequency estimation using periodic projective measurements, for handy comparison to the continuous case.

\section{Static Parameter Estimation}\label{sec:staticpe}
We start by describing the problem. Generally speaking, we consider the estimation of an unknown fixed parameter in the system Hamiltonian, given the output of some quantum measurement device. Suppose, for definiteness, that we are interested in measuring the Rabi oscillation rate of a driven qubit from the measurement output data. The unmeasured qubit is then described by a Hamiltonian
\be 
H= \frac{\hbar \Omega}{2}\, Y \label{ham},
\ee
that rotates the qubit state in the X-Z plane of the Bloch sphere at an angular frequency $\Omega$, where we notate the usual Pauli operators as $\{X,Y,Z\}$. In the absence of measurement, the quantum system unitarily evolves for a duration $\delta t$, which can be simply represented as a rotation in the configuration basis,
\be
U = \begin{pmatrix} \cos \frac{\Omega \,\delta t}{2} & -\sin \frac{\Omega \,\delta t}{2} \\ \sin \frac{\Omega \,\delta t}{2} & \cos \frac{\Omega \,\delta t}{2}
\end{pmatrix}. \label{u}
\ee
We now wish to estimate the oscillation frequency $\Omega$. 

We begin our analysis by assuming a time-independent oscillation frequency $\Omega$ that is unknown beforehand, and we examine different methods for estimating the parameter. In Sec.~\ref{sec:maxlike}, we consider a maximum likelihood-based matrix multiplication method to estimate the parameter of interest, showing both idealized and more realistic cases that include experimental nonidealities. In Sec.~\ref{sec:fourier}, we compare this method  to a simpler method based on FFT, and show that the FFT permits a quick-but-crude estimation of a prior range of frequencies that can be used to help accelerate the convergence of the maximum likelihood procedure.

\subsection{Maximum Likelihood}\label{sec:maxlike}
Maximum likelihood methods presuppose a model that describes the stochastic physics with fixed parameters, then varies each unknown parameter to find the best estimate matching a target data set according to a suitable likelihood measure. In our case, the model is given by the quantum Bayesian update corresponding to repeated unsharp (generalized) measurements applied to a single qubit undergoing Hamiltonian evolution. The likelihood measure is simply the probability for obtaining the observed sequence of measurement results.

\subsubsection{Ideal Continuous Measurement}
The physical set-up of the Hamiltonian of the system is described by Eq.~\eqref{ham}, to which we now add a (Markovian) continuous measurement of the $Z$ operator. A thorough analysis of such a system may be found in Ref.~\cite{koro01}, which considers a solid-state qubit formed by a double-quantum-dot and measured by the current flowing through a nearby quantum point contact. Notably, the analysis found therein applies mostly unchanged to more recent superconducting qubit measurements \cite{vija11} that use circuit quantum electrodynamics \cite{gamb08,koro11}. We consider an idealized such model here, to which we will later add additional experimental imperfections.

Such a continuous measurement weakly probes information about the qubit state coordinate $z(t) \equiv \mean{Z}(t)$ from the system, producing a (suitably renormalized) noisy record $r(t) \approx z(t) + \sqrt{\tau_m}\,\xi(t)$ that is approximately centered around $z(t)$ and masked with Gaussian white noise $\xi(t)$ (satisfying the correlation $\mean{\xi(0)\xi(t)} = \delta(t)$). The \emph{characteristic measurement time-scale} $\tau_m$ in the noise-power determines the amount of time needed to distinguish the eigenstates $Z = \pm 1$ with unit signal-to-noise ratio \cite{koro11}. 

In practice, such continuous readouts $r(t)$ are digitized by the hardware into time bins $t_j \equiv j\,\delta t$ of duration $\delta t$. The output reported by the detector is thus a time-sliced picture of discrete outputs $r_j = \int_{t_j}^{t_{j+1}}r(t)\,dt/\delta t$, that are temporal averages of the continuous signal $r(t)$ over each time bin $t_j$. After a total duration $T \equiv N\,\delta t$, a measurement read-out $\{r_j\}$ is thus produced, consisting of real outputs $r_j$ at each time step $j=1, \ldots, N$. Our goal is to directly use such a readout $\{r_j\}$ to estimate the Rabi oscillation frequency $\Omega$ in Eq.~\eqref{ham}.

We use an effective (\emph{quantum Bayesian}) measurement model for such a time-sliced $Z$ measurement (see, e.g., Ref.~\cite{jordan2010uncollapsing,koro11}), that models the readout $r$ for each time slice $t_j$ as an independent random variable sampled from a Gaussian mixture distribution,
\begin{eqnarray}\label{eq-readout}
P(r) &=& \rho_{11}\, P(r|1) + \rho_{00}\, P(r|0), \\
\label{eq-readout2}
P(r|0, 1) &=& \sqrt{\frac{\delta t}{2\pi \tau_m}} \exp\left[- \frac{\delta t(r \pm 1)^2 }{2\tau_m}\right],
\end{eqnarray}
with $P(r|0)$ centered on $r=-1$ and $P(r|1)$ centered on $r=+1$, and where the probabilities $\rho_{11}$ and $\rho_{00}$ are density-matrix elements that correspond to the qubit $Z$-populations at the beginning of the time-slice $t_j$. For sufficiently short time-slices $\delta t$, this Gaussian mixture approximates a single broad Gaussian of variance $\tau_m/\delta t$ that is centered at the qubit coordinate $z \equiv \rho_{11} - \rho_{00}$, thus recovering the Gaussian white noise picture $r(t) \approx z(t) + \sqrt{\tau_m}\,\xi(t)$ in the continuum limit.

For each time-step $t_j$ with duration $\delta t \ll \tau_m$, the qubit is only weakly perturbed by the measurement while it evolves with the Hamiltonian in Eq.~\eqref{ham}. The state backaction between $t_j$ and $t_j + \delta t$ thus combines evolution with a partial-collapse of the prior system state $\rho$ at $t_j$, and is described by the update rule
\be \label{eq-rhoM}
\rho' = \frac{M_r \rho M_r^\dagger}{Tr[M_r^\dagger M_r \rho]},
\ee
in terms of \emph{measurement operators} $M_r$ that depend upon the readout $r = r_j$ observed in the interval $[t_j,t_{j+1}]$. For time steps $\delta t \ll (2\pi/\Omega)$ much smaller than a Rabi period, each measurement operator $M_r \approx U E_r^{1/2}$ can be approximately decomposed into a unitary part $U$ and a positive operator $E_r$, such that $U$ is given by the ($r$-\emph{in}dependent, $\Omega$-dependent) unitary evolution of Eq.~(\ref{u}) over the elapsed time $\delta t$, and the positive operators $E_r$ are ($r$-dependent, $\Omega$-\emph{in}dependent) elements of a \emph{positive operator-valued measure} (POVM) that is fully determined by the Gaussian probabilities in Eq.~\eqref{eq-readout2}
\be\label{eq:povm}
E_r = \begin{pmatrix} P(r|0) & 0 \\ 0 & P(r|1)
\end{pmatrix}.
\ee
Note that full collapse of the wavefunction would be obtained by taking $\tau_m \to 0$ (Zeno measurement regime) \cite{nielsen2010quantum}, in which case $E_r$ would converge to projection operators for each definite $Z$-state of the qubit. Importantly, the probability of having obtained the readout $r$ given the prior state $\rho$ is then $P(r|\rho) = {\rm Tr}[M_r\rho M_r^\dagger] = {\rm Tr}[E_r\,\rho]$, which reproduces Eq.~\eqref{eq-readout}.

\subsubsection{Maximum Likelihood Estimation}
For the estimation of the unknown value of $\Omega$ from the record taken over a total duration $T = N\,\delta t$, we require the joint probability distribution for a long sequence of results $\{ r_1, r_2, \ldots, r_N\}$, which according to Eq.~\eqref{eq-rhoM} has the simple form
\be
P(r_1, \ldots, r_N|\Omega) = {\rm Tr}[M_{N} \rho M_{N}^\dagger], \label{dist}
\ee
where $\rho$ is the known initial state and where we have defined an effective, multi-index measurement operator,
\be
M_{N}(r_1, \ldots, r_N) = M_{r_N} \ldots M_{r_2} M_{r_1}
\ee
as the simple product of the measurement operators $M_{r_j}$ for each result $r_j$.

Given no prior information about the value of $\Omega$, we use the log of the distribution in Eq.~\eqref{dist} as our {\it log-likelihood function} for an observed readout
\be\label{eq:loglike}
{\cal L}(\Omega) = \ln P(r_1,\ldots,r_N|\Omega) = \ln {\rm Tr} [M_N^{\dagger}M_N \rho].
\ee
That is, given an observed data set $(r_1, \ldots, r_N)$ from an experiment, we are interested in finding the {\it maximum likelihood estimator} $\Omega_{ML}$, such that 
the distribution in Eq.~\eqref{dist}, and therefore Eq.~\eqref{eq:loglike}, is maximized for $\Omega = \Omega_{ML}$. Importantly, it is \emph{not} necessary to track the quantum state (e.g., by using Eq.~\eqref{eq-rhoM} or solving a stochastic mater equation), since we are only interested in estimating the parameter $\Omega$.  

Consequently, only the functional form of the measurement operators are needed, and only the state-dependent part of each $M_{r_j}$ will be relevant for the MLE optimization. As such, the state-independent Gaussian normalization may be conveniently discarded as uninformative. In what follows we rescale the measurement operator $M_{r_j} \mapsto U\tilde{E}_{r_j}^{1/2}$ by using an unnormalized equivalent of Eq.~\eqref{eq:povm} that isolates the state-dependent part 
\be
{\tilde E}_j = \begin{pmatrix} \exp(- r_j \delta t/\tau_m) & 0 \\ 0 & \exp(r_j \delta t/\tau_m).
\end{pmatrix}.
\ee
This replacement will later help numerical algorithms avoid products of very small numbers.

The Fisher information about the parameter $\Omega$ to be estimated is computed directly from the log-likelihood for all possible readouts 
\be
{\cal I}(\Omega) = \int\!\! {\cal D}r P(r_1, \ldots r_N) \left( \partial_\Omega \ln P(r_1, \ldots, r_N|\Omega)\right)^2,
\ee
where the measure ${\cal D}r$ indicates integration over all possible values of the $N$ measurement results $r_j$. The Fisher information determines the minimum variance of the parameter $\Omega$, the Cram\'er-Rao bound,
\be
{\rm Var}\, \Omega \ge {\cal I}^{-1},
\ee
As we show in the Appendix for a related special case using periodic projective measurements, this bound on the variance can be saturated by choosing the maximum likelihood estimation (MLE) procedure.

The uncertainty $\sigma$ of the estimate that maximizes the log-likelihood function and saturates the Cram\'er-Rao bound is given by the observed Fisher Information itself,
\be\label{eq:uncertainty}
\sigma^{-2} = -\partial_\Omega^2 \ln P|_{\Omega = \Omega_{ML}}.
\ee
which can be conveniently found as the width of a parabolic fit to the computed log-likelihood function around the maximum $\Omega_{ML}$, ${\cal L}(\Omega) \approx -(\Omega-\Omega_{ML})^2/2\sigma^2$.

From this point, there are two ways to proceed.
The first semi-analytic way is to examine the maximum likelihood condition, $\partial_\Omega {\cal L}(\Omega)=0$, given the (measured) data set $\{ r_j\}$, which then yields the equation
\be
{\rm Tr} [((\partial_\Omega M_N^\dagger) M_N + M_N^\dagger \partial_\Omega M_N) \rho]=0.
\ee
This equation must be solved numerically in general. An iterative procedure is also possible since the measurement operator is a product of all previous operators. Therefore, we have the recursion relation
\be
\partial_\Omega M_N = (\partial_\Omega M_{r_N}) M_{N-1} + M_{r_N} (\partial_\Omega M_{N-1}),
\ee
where $M_1 \equiv M_{r_1}$, and where $\partial_\Omega M_{r_j} = (\partial_\Omega U)\tilde{E}_{r_j}^{1/2}$ depend on $\Omega$ only through the unitary $U$. That is, to compute an updated likelihood from the likelihood at the previous step, we can use the already computed $M_{N-1}$ and its derivative, as well as the newly computed measurement matrix $M_{r_N}$ and its derivative to proceed to the next time step. This recursive simplification permits efficient parallel calculation of likelihood values for a range of $\Omega$ values during data collection.

The second (simpler) way is to entirely numerically implement the MLE method. For each result $r_j$ in the string of incoming data $( r_1, r_2, \ldots, r_N )$ we numerically compute the relevant matrix $M_{r_j} = U\tilde{E}_{r_j}^{1/2}$ that includes one of a variety of (precomputed) unitary matrices $U$ that assume possible $\Omega$ values, as well as the rescaled POVM element $\tilde{E}_{r_j}$. For each chosen $\Omega$ value, we then compute the measurement operator $M_N$ as a product of the string of $N$ matrices $M_{r_j}$, and compute its associated log-likelihood function. 
This procedure results in a discretized function ${\cal L}(\Omega)$ of log-likelihoods over all sampled $\Omega$. The best estimate $\Omega_{ML}$ maximizes this function, while the uncertainty $\sigma$ is the best quadratic fit around this maximum according to Eq.~\eqref{eq:uncertainty}. 

\begin{figure}[t]
\includegraphics[width=\columnwidth]{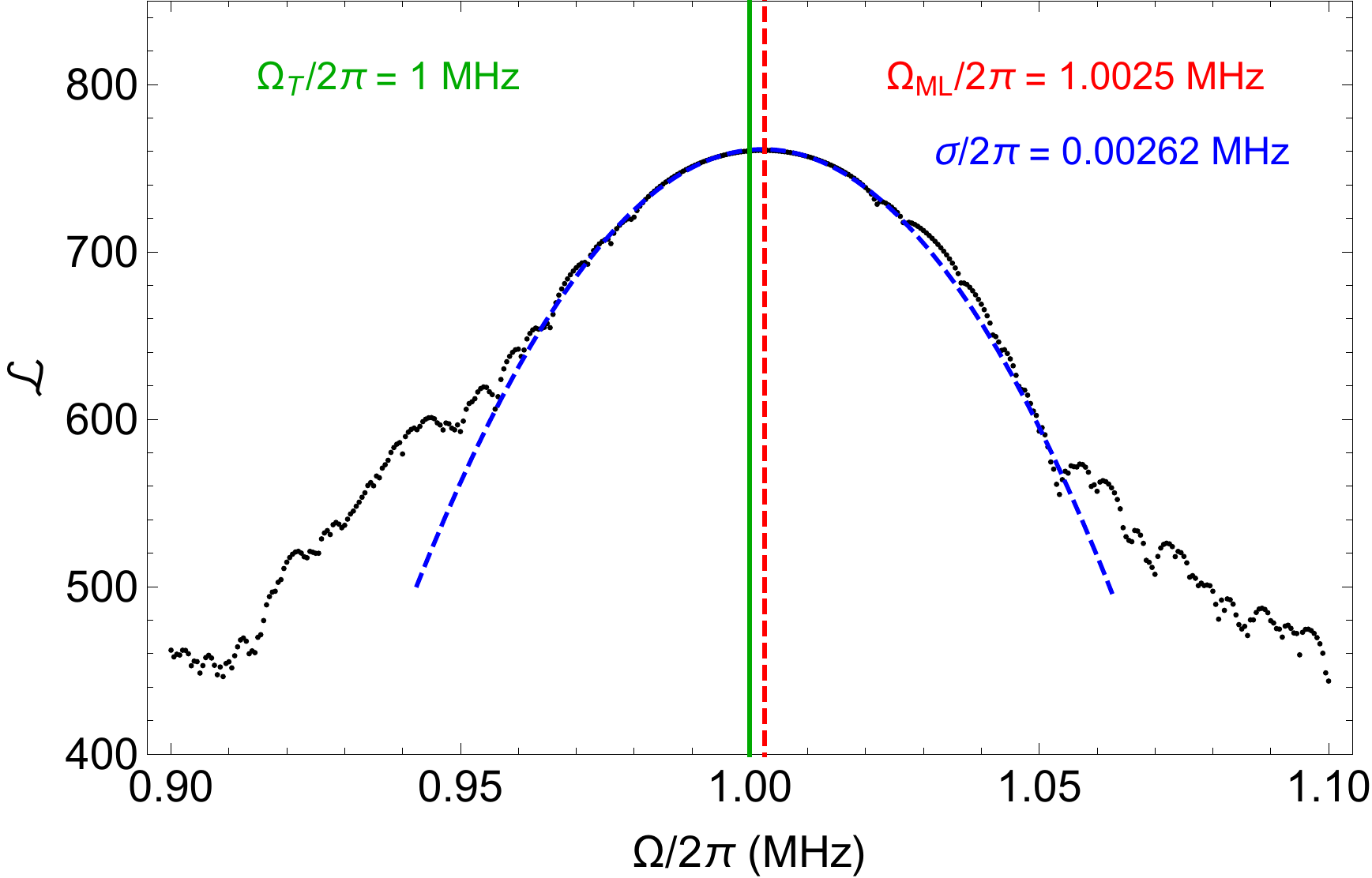}
\caption{Log-likelihood ${\cal L}(\Omega)$ as a function of possible Rabi frequencies $\Omega$. A long noisy $Z$-measurement record $\{r_j\}$ of duration $T = 1$ms was simulated with $N = 10^5$ discrete time steps $\delta t = 10$ns, a characteristic measurement time $\tau_m = 1 \mu$s, and a true Rabi oscillation frequency $\Omega_T/2\pi = 1$ MHz (green solid). The maximum likelihood estimator (red dotted) is the peak at $\Omega_{ML}/2\pi = 1.0025$ MHz with $0.25\%$ error. The quadratic fit ${\cal L} \approx -(\Omega - \Omega_{ML})^2/2\sigma^2$ to the peak yields the precision $\sigma/2\pi = 0.0026$ MHz.}
\label{fig-loglike}
\end{figure}

\subsubsection{Numerical Simulations}
Although frequency is a continuous parameter, the numerical maximum likelihood algorithm outlined above must search over a discrete set of trial values for the frequency. An initial search grid coarse-grains a frequency range of interest in units of $\delta \Omega$, and computes the maximum likelihood over this grid. Both the precision of the estimation and the computational efficiency of the procedure thus depend directly on the size of the search grid. A coarse frequency step size defines a big grid, which makes the search faster but with lower precision, and vice versa. In general, an adaptive mesh size is useful to find the desired precision, where the spacing between sampled $\Omega$ is iteratively refined to increase the resolution around the maximum.

As an example of such a purely numerical maximum likelihood method, we simulate a single measurement readout with a chosen true value of the Rabi oscilation frequency $\Omega_T=2\pi f$ with $f=1$ MHz, while monitoring with a characteristic measurement time $\tau_m = 1\mu$s at discete time steps $\delta t = 10$ns for a duration $T = 1$ms (${\sim}$1000 oscillations), and then compute ${\cal L}$ for a range of frequencies around $\Omega_T$. We plot the (unnormalized) log-likelihood function in Figure~\ref{fig-loglike}, where a dominant peak is clearly seen around the correct value. An estimation of the precision is given by fitting the log-likelihood with a polynomial function (dashed-blue), $- (\Omega - \Omega_{ML})^2/2 \sigma^2$, around the peak $\Omega_{ML}/2\pi$. The estimated value of the frequency is $\Omega_{ML}/2\pi = 1.0025$ MHz, which shows 0.0025 MHz ($0.25\%$) error with a precision of $\sigma/2\pi = 0.0026$ MHz, which is close to the minimum anticipated frequency resolution of $1/T = 0.001$ MHz.


The proof-of-principle simulation shown above demonstrates that MLE allows us to determine the oscillation frequency $\Omega$ quite precisely using a measurement record about 1000 times the length of the characteristic measurement time $\tau_m$, over a span of roughly 1000 oscillations.  However, realistic experiments have other characteristic timescales (like energy relaxation and dephasing) that will practically bound the duration $T$ over which a measurement may be taken. Moreover, the frequency $\Omega$ itself may exhibit slow drift over longer timescales, which again practically bounds the duration $T$. We will consider both these cases later in this paper.

Keeping these practical limitations in mind for now, it is advantageous to optimize the estimation to use shorter measurement records. The estimation error will depend on two free timescale parameters, the duration $T$ and the measurement time $\tau_m$, in addition to a fixed timescale, the true Rabi period $2\pi/\Omega_T$. Hence, for a given target estimation error we wish to minimize the duration $T$ by optimizing the measurement time $\tau_m$.

\begin{figure}
\includegraphics[width=\columnwidth]{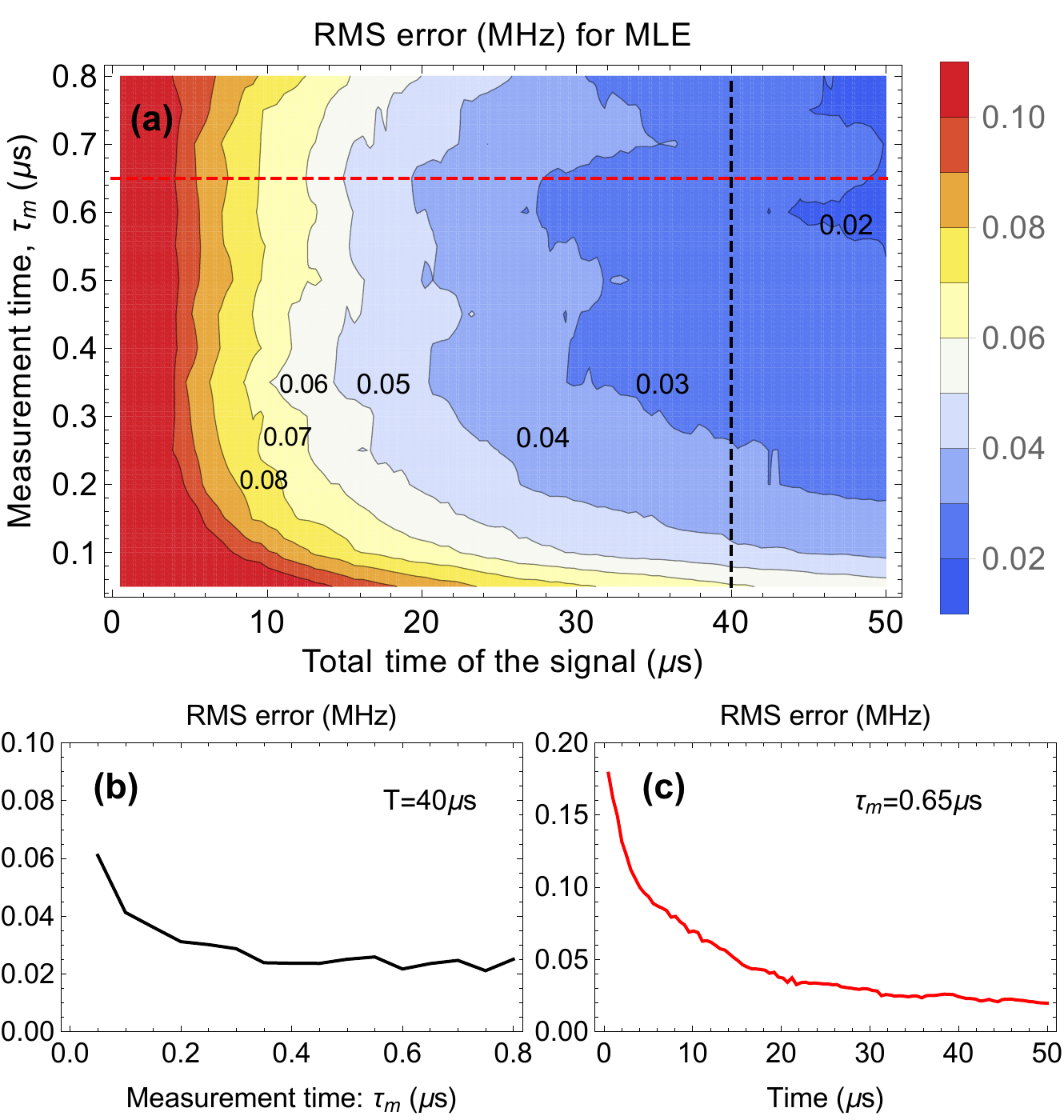}
\caption{RMS error of maximum likelihood estimation (MLE) of frequency versus measurement time $\tau_m$ and total signal duration $T$, with $N_E = 600$ trajectory realizations. (a) Error of MLE, indicating that longer signals yield more accurate estimates, for a range of optimal measurement times $\tau_m$.  (b) Slice of (a) with constant $T = 40\,\mu$s, showing the ``sweet spot'' where the RMS error does not strongly depend on $\tau_m$.  (c) Slice of (a) with constant $\tau_m = 0.65\,\mu$s, showing the reduction of error with increased collection time.}
\label{contourMLE}
\end{figure}

To determine how the estimation error behaves as both $T$ and $\tau_m$ are varied for a fixed $\Omega_T$, we numerically simulated the error for a range of parameters. The duration $T$ of each simulation was varied between $T=1$--50 $\mu$s in increments of $1\,\mu$s. For each $T$, the measurement time $\tau_m$ was also varied between $\tau_m=0.05$--$0.8$ $\mu$s in increments of $0.05\,\mu$s. The root-mean-square (RMS) deviation from the true frequency $\Omega_T$  
\be
\Omega_{RMS}=\sqrt{\frac{\sum_i (\Omega_i-\Omega_T)^2}{N_E}},
\ee
was computed over an ensemble of $N_E=600$ realizations for each parameter choice $(T,\tau_m)$ to quantify the estimation error. Results for the statistics are shown in Fig.~\ref{contourMLE}(a) as a contour plot. Fig.~\ref{contourMLE}(c) shows the horizontal slice through the dashed line corresponding to the optimal measurement time of $\tau_m = 0.65\,\mu$s, and shows the improvement of the estimation error (${\sim}1/\sqrt{T}$) with an increase in $T$. Fig.~\ref{contourMLE}(b) shows the vertical slice through the dashed line corresponding to $T = 40\,\mu$s, showing that there is an optimum $\tau_m$ that minimizes the error when $T$ is held fixed.

\subsubsection{Including Non-idealities}
\label{sec:non-idealities}
The above discussion assumed that the stochastic evolution from the continuous measurement preserved the purity of the state. However, more realistic evolution must include additional experimental nonidealities that decrease the state purity. These nonidealities include qubit energy-relaxation with a characteristic timescale $T_1$, environmental dephasing of the qubit to its energy basis with a characteristic timescale $T_2$, and collection loss within the readout chain that leads to a net collection efficiency $\eta\in[0,1]$.  Other unexpected environmental effects over longer timescales that cause frequency drift (such as thermal fluctuations) will be considered later.

Including these effects requires an extended model of the state dynamics (and thus the maximum likelihood method) from pure states to mixed states. A particularly useful representation that accommodates mixed states is the Bloch paravector picture: given an unnormalized density operator $\rho$, such a paravector has the 4 real state coordinates $\vec{\rho} \equiv (x,y,z,p)$, such that $x = {\rm Tr}[X\rho]$, $y = {\rm Tr}[Y\rho]$, $z = {\rm Tr}[Z\rho]$, and $p = {\rm Tr}[\rho]$. Here the state normalization $p$ physically indicates the probability that the state was prepared. Dividing by this $p$ renormalizes the state $\vec{\rho}\mapsto (x/p,y/p,z/p,1)$ so that its first three coordinates are the usual Bloch coordinates (expectation values) conditioned on definite (successful) state preparation. By including the normalization of the state explicitly, we can linearize the state evolution due to measurement into a matrix product that generalizes the pure state case of Eq.~\eqref{eq-rhoM}.

Specifically, the ideal case of $\rho \mapsto M_r \rho M_r^\dagger$ with $M_r = U\tilde{E}^{1/2}_r$ can be written equivalently as the $4\times 4$ matrix product $\vec{\rho} \mapsto \mathbf{M}_r\vec{\rho}$, where $\mathbf{M}_r = \mathbf{V}\mathbf{F}_r$ and 
\begin{align}
\mathbf{V} &\equiv \begin{bmatrix}\cos(\Omega\, \delta t) & -\sin(\Omega\,\delta t) & 0 & 0 \\\sin(\Omega\,\delta t) & \cos(\Omega\,\delta t) & 0 & 0\\0 & 0 & 1 & 0\\0 & 0 & 0 & 1\end{bmatrix}, \\
\mathbf{F}_r &\equiv \begin{bmatrix}1 & 0 & 0 & 0 \\ 0 & 1 & 0 & 0 \\ 0 & 0 & \cosh(r\,\delta t/\tau_m) & \sinh(r\,\delta t/\tau_m) \\ 0 & 0 & \sinh(r\,\delta t/\tau_m) & \cosh(r\,\delta t/\tau_m)\end{bmatrix}.
\end{align}
Notably, over the duration $\delta t$ the $Z$-measurement acts as a hyperbolic rotation in the $z$-$p$ plane, while the Hamiltonian evolution acts as an elliptic rotation in the $x$-$y$ plane. Interestingly, this behavior is completely analogous to the boosts and rotations of spacetime coordinates in Lorentz transformations \cite{jordan2006qubit}. To add nonidealities, it is convenient to re-express these rotations in terms of their infinitesimal generators, which are identical to the Lorentz transformation generators 
\begin{align}
  \mathbf{V} &\equiv \exp\left(\delta t\begin{bmatrix}0 & -\Omega & 0 & 0 \\ \Omega & 0 & 0 & 0 \\ 0 & 0 & 0 & 0 \\ 0 & 0 & 0 & 0\end{bmatrix}\right), \\
  \mathbf{F}_r &\equiv \exp\left(\delta t\begin{bmatrix}0 & 0 & 0 & 0\\ 0 & 0 & 0 & 0 \\ 0 & 0 & 0 & r/\tau_m \\ 0 & 0 & r/\tau_m & 0\end{bmatrix}\right).
\end{align}
Properly, these generators should be summed before the exponentiation over the interval $\delta t$, but we keep them separated here for conceptual clarity and simplicity in the maximum likelihood calculation. Note that these expressions make it clear that both $\Omega$ and the averaged stochastic result $r$ are assumed constant over the discretization interval $\delta t$.

We can now readily add three types of nonideality.  First, we include collection inefficiency $\eta\in[0,1]$, such that the dephasing rate of the qubit due to measurement is $\Gamma_m \equiv 1/(2\eta\tau_m)$. This definition implies that the timescale $\tau_m$ indicates the actual information acquisition observed at the detector, while the measurement dephasing rate $\Gamma_m$ of the qubit also includes the averaged backaction of the signal that was never detected \cite{koro11}. Second, we include environmental dephasing with timescale $T_2$ that arises from sources other than measurement, which contributes an extra term $1/T_2$ to the total qubit dephasing rate. Third, we add energy-relaxation of the qubit at the timescale $T_1$. This type of nonideality is more complicated; it preserves the probability $p$ while shifting the population toward the ground state at an exponential rate $1/T_1$. In the $z$-$p$ coordinates, this has the form, $z\mapsto z\,e^{-\delta t/T_1} - p(1-e^{-\delta t/T_1})$. This shift in population also results in an additional dephasing term $(x,y) \mapsto (x,y)e^{-\delta t/2T_1}$. Writing all three of these nonidealities using their infinitesimal generators, we obtain
\begin{align}
  \mathbf{F}_r &\equiv \exp\left(\delta t\begin{bmatrix}-\gamma & 0 & 0 & 0\\ 0 & -\gamma & 0 & 0 \\ 0 & 0 & -1/T_1 & r/\tau_m - 1/T_1 \\ 0 & 0 & r/\tau_m & 0\end{bmatrix}\right),
\end{align}
in terms of the total dephasing rate
\begin{align}
  \gamma &\equiv \Gamma_m + \frac{1}{T_2} + \frac{1}{2T_1}, & \Gamma_m &\equiv \frac{1}{2\eta\tau_m}.
\end{align}
This matrix exponential can easily be written in a closed form (and should be for numerical efficiency), but we omit it here for brevity. It is also worth recalling that this form of the measurement-evolution neglects normalization factors to simplify the linear evolution---the state may always be renormalized after each step during simulation if desired according to $\vec{\rho} \mapsto \mathbf{M}_r\vec{\rho} / (\vec{1}\cdot\mathbf{M}_r\vec{\rho})$, where $\vec{1} \equiv (0,0,0,1)$ extracts the new state norm.

We can then write the log-likelihood function as
\begin{subequations}
\begin{align}
  {\cal L}(\Omega) &\equiv \ln \left(\vec{1} \cdot \mathbf{M}_N\vec{\rho}\right), \\
  \mathbf{M}_N &= \mathbf{M}_{r_N}\cdots\mathbf{M}_{r_1}.
\end{align}
\end{subequations}
The product of $2\times 2$ matrices $M_N$ from before has been simply replaced with the product of $4\times 4$ matrices $\mathbf{M}_N$, while the pure state has been generalized to a mixed state vector. With this simple change, the rest of the maximum likelihood procedure outlined in the previous sections proceeds unaltered. 

\subsection{Fourier Methods}\label{sec:fourier}

As we have shown, MLE methods can be used to accurately estimate the drive frequency. However, 
the necessary time to run the algorithm
depends crucially on the frequency search range in the MLE algorithm. In order to speed up this search it is beneficial to have at least a rough estimate for the value of $\Omega_T$.

\begin{figure}[t]
\includegraphics[width=\columnwidth]{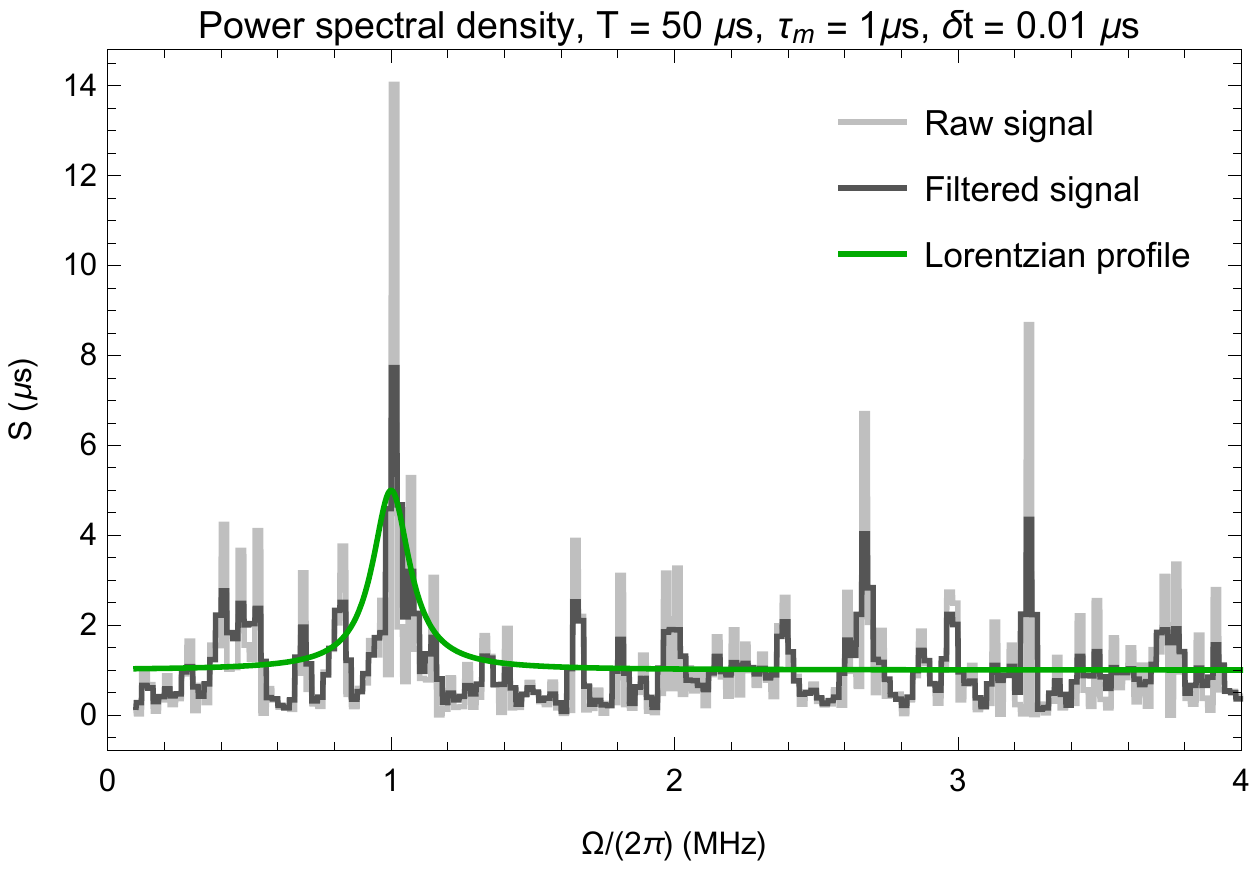}
\caption{Power spectral density for $T = 50\, \mu$s, $\tau_m = 1\, \mu$s and $\delta t = 0.01\, \mu$s, shown for a single realization of the measurement process (gray). After applying a triangular moving-average filter (black) with a width of roughly $T/(2\pi \tau_m)$ points of minimum frequency resolution $\delta \Omega/2\pi = 1/T$, the fluctuations in the power spectral density are reduced. With sufficient nearest-neighbor averaging, the filtered data will approach a Lorentzian profile (green) with increasing $T$.
}
\label{fig:LorentzianComparison}
\end{figure}

We know the MLE method should saturate the Cram\`er-Rao bound, so any other estimation method should perform less well. Nevertheless, other methods may have other desirable qualities, such as estimation speed. A simple and natural way to coarsely-yet-quickly estimate the drive frequency is by studying the power spectral density $S(\Omega)$ of the output signal $r(t)$, which can be quickly estimated from the fast Fourier transform (FFT) of the discretized signal ($\{r_j\}$ at time points $\{t_j\}$ with spacing $\delta t$) as $S(\Omega_j) = \big|\mathrm{FFT}_{t_j\to\Omega_j}\left[\{r_j\}\right]\big|^2 \delta t$, where the resulting discrete frequencies $\{f_j \equiv \Omega_j/2\pi\}$ have spacing $\delta f = 1/T = 1/(N \delta t)$, where $N$ is the number of time steps in the original signal $\{r_j\}$.
Even a plain FFT method is enough to accurately estimate the frequency for some regime of parameters. However, better estimates can 
be achieved by simple filtering methods. 
Fig.~\ref{fig:LorentzianComparison} shows a simulated power spectral density of the output signal for $\tau_m = 1\, \mu$s, a total runtime of $T = 50\, \mu$s, and $\delta t = 0.01\, \mu$s, compared to a filtered power spectral density (using a triangular center-weighted moving average over 5 nearest bins of width $\delta f$).
As the figure illustrates, in this weak measurement regime the spectral density tends to a Lorentzian function as the total runtime increases~\cite{Korotkov01}. 
The background noise is concentrated around $\tau_m$, with a peak centered around the drive frequency of height $4\tau_m$. 
Moreover, the measurement time also characterizes the width of the peak, with a Full-Width-Half-Maximum of $1/(2 \pi \tau_m)$. 
The total runtime $T$ and time step $\delta t$ determine the resolution and maximum frequency of the Fourier transform method, with $\delta f = 1/T$ and $f_{\text{max}} = 1/ \delta t$, respectively.
The resolution $\delta f$ dictates the maximum size of the filtering window that one can apply, given that roughly $1/(2\pi \tau_m \delta f) = T/(2\pi \tau_m)$ points fit inside the peak of the Lorentzian.

\begin{figure}[t]
\includegraphics[width=\columnwidth]{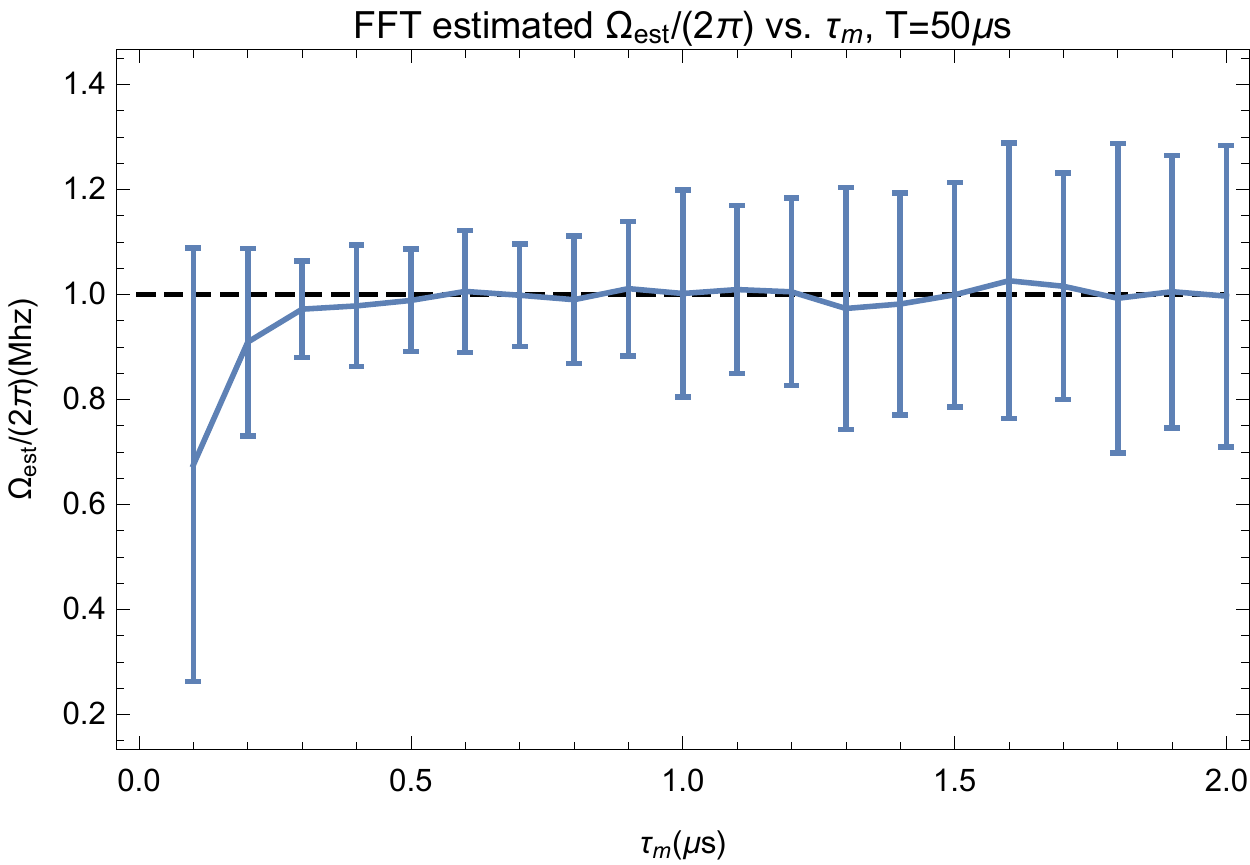}
\caption{Frequency estimation via fast Fourier transform. 
The estimation of the peak frequency can be optimized by filtering the spectral density, as shown in Fig.~\ref{fig:LorentzianComparison}. For a total runtime of $T = 50\, \mu$s and time step $\delta t = 0.01\, \mu$s the estimated frequency approaches the real frequency $\Omega_T/2\pi = 1$ MHz. The RMS error averaged over $N_E=200$ realizations is of the order of $10\%$ for measurement times $\tau_m \sim 0.3 \mu$s to $\tau_m \sim 0.8 \mu$s. However, this error increases for shorter measurement times due to zero frequency Zeno pinning, or for longer measurement times, where the measurement is too weak for the output signal to accurately estimate the frequency.
}
\label{fig:FFTlong}
\end{figure}

In Fig.~\ref{fig:FFTlong} we show the RMS error of the FFT estimated frequency after filtering for 200 realizations of the measurement, as a function of the measurement time $\tau_m$, for a total runtime $T = 50\, \mu$s. In this parameter regime the simple FFT method is fairly effective at estimating the frequency to within $10 \%$ error for $\tau_m$ between $0.3\, \mu$s to $0.8\, \mu$s.
Notice, however, that as $\tau_m$ decreases the measurement process tends to `pin' the system into one of the eigenstates of the $Z$ operator. Hence, in this ``Zeno regime'' the spectral density shows a peak at zero frequency \cite{koro01}, hindering the FFT frequency estimate and increasing the average error.

\begin{figure}[t]
\includegraphics[width=\columnwidth]{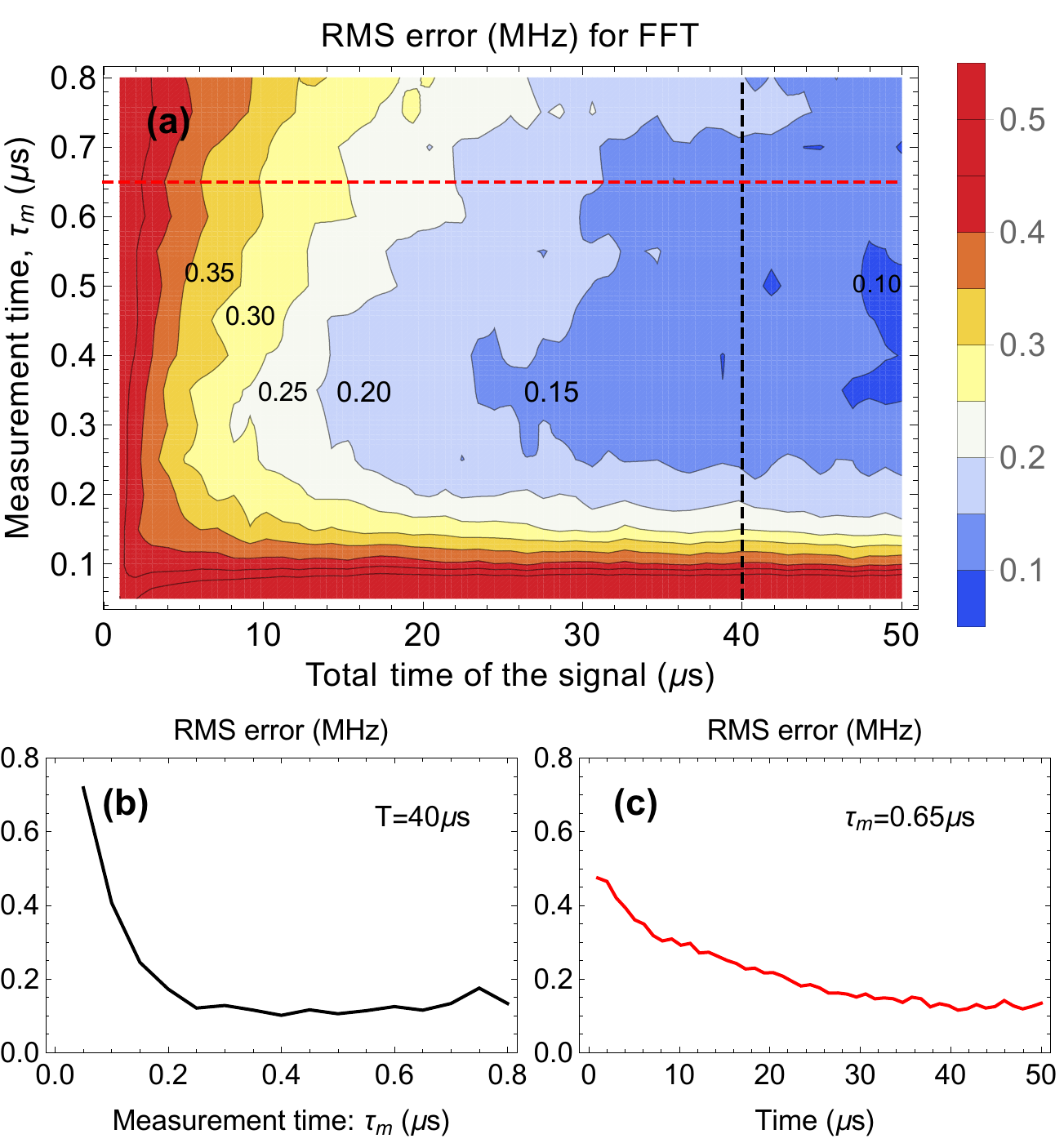}
\caption{RMS error of fast Fourier transform (FFT) frequency estimate versus measurement time $\tau_m$ and total signal duration $T$, with $N_E = 600$ trajectory realizations, and a bandpass filter to a window $\Omega\in[0,\,2\Omega_T]$, as well as a triangular moving average over nearest spectral points of width $\delta \Omega/2\pi = 1/T$. (a) Error of FFT, showing uniformly worse performance than MLE.  (b) Slice of (a) with the same constant $T = 40\,\mu$s as Fig.~\ref{contourMLE}. (c) Slice of (a) with the same constant $\tau_m = 0.65\,\mu$s. These crude estimates may be used to accelerate MLE. } \label{fig:FFT_contour}
\end{figure}

Since, as we saw in the previous section, the MLE approach is optimal, we expect that it will outperform the Fourier transform in estimating the drive frequency with the same duration $T$. In Fig.~\ref{fig:FFT_contour}, we show how the RMS error scales with both runtime $T$ and measurement time $\tau_m$ using the FFT method, which should be compared with the MLE scaling in Fig.~\ref{contourMLE} of the previous section. The best FFT error for short times ranges from $10$--$20$\% error, compared with $2$--$5$\% error obtained by MLE. The deterioration of performance for small $\tau_m$ due to Zeno pinning is also clearly visible. Note that for sake of comparison with MLE we have used a bandpass filter of width $2$~MHz centered around $\Omega_T/2\pi = 1$~MHz in Fig.~\ref{fig:FFT_contour} to eliminate false positives at higher frequencies. For much longer durations $T$, the power spectral density produced by the FFT may be more aggressively window-averaged, so the bandpass filter may be removed.

\subsection{Computational Efficiency}\label{sec:efficiency}
To see when FFT will provide an improvement over MLE in terms of computational efficiency, let us compute a rough estimate of how each method scales with the number of time points $N = T/\delta t$.

For the FFT method, the dominant contribution to the computation is performing the FFT itself, which scales as $N\log{N}$, with the other operations (squaring, filtering, maximizing) scaling linearly with $N$. For our improved MLE method, with a grid of $n$ trial frequencies and matrices of dimension $d$, the dominant contribution scales as $d^3\,n\, N$ due to matrix multiplications within a loop over trial frequencies, with other contributions (trace, maximization) scaling at most linearly in $N$. For our method, $d$ is either 2 or 4, thus giving an overall constant prefactor. The relevant comparison for the dominant scaling is thus $N\log{N}$ for FFT vs. $n\,N$ for MLE.

Using an initial long duration $T$, the computational efficiency of the FFT method 
may be used to accelerate the MLE estimation. The FFT-estimated frequency can then be used as starting point for the more precise MLE algorithm using shorter runtimes $T$. 
We note the the unitaries $U_j$ (depending on the trial $\Omega_j$) can be pre-computed as numerical matrices, and simply called from a database when the given trial frequencies appear in order to optimize the algorithm.  If one does not have any prior information about the possible value of the drive frequency, then one needs a large grid with $n \sim N$ points, so FFT will be faster than MLE by a crude factor of $\log{N}/N$. However, after using an initial FFT estimation to reduce the set of trial frequencies, then MLE may use a significantly smaller grid. If the grid is made sufficiently small, such that $n < \log{N}$, then MLE will be faster than FFT so that it can be appended with overhead scaling subdominantly in $N$.  

As we will explore in detail in the next section, the scaling of MLE makes it particularly well suited for dynamical estimation problems. In such a problem, the frequency may be initially estimated via a combined FFT and MLE method, followed by periodic updates of this estimate using a faster MLE estimator scaling only linearly in $N$.

\section{Dynamic Parameter Estimation}\label{sec:dynpe}
The preceding analysis involves the estimation of a fixed Rabi frequency $\Omega$. We now consider the following (more interesting) problem: what if the Rabi frequency $\Omega$ is not fixed, but instead changes in time due to experimental drift caused by changes in the environmental or control system at longer timescales (e.g., thermal variation)? We would then like to update our estimate $\Omega(t)$ continuously in time, which would allow us to monitor and compensate for such a drift \emph{in situ}, in near-real-time. To accomplish this goal, we consider a moving window of a fixed duration $T$, such that the end-point of this window is the current time $t$ (i.e., a fixed-delay scenario). Estimating the frequency as before within such a moving window then produces a time-evolving maximum-likelihood estimate $\Omega_{ML}(t)$. There is thus a trade-off between the maximum precision of the frequency estimation allowed by the chosen window ($\sigma > \delta f = 1/T$), and the resulting sensitivity to the timescale of a detectable drift $T_d > T$, which should be longer than the temporal averaging resulting from the estimation window.

As discussed in Sec.~\ref{sec:efficiency}, by using an initial estimation (either FFT or MLE, or a combination) to narrow the frequency search range---assuming the drift is slow compared to the time scale of the estimation---the MLE algorithm scales only linearly in the number of time steps $N$, making real-time updates to the initial estimation computationally reasonable.

\subsection{Including prior information in the estimate}
In contrast to Sec.~\ref{sec:maxlike}, where we assumed that we had no prior information about the parameter we are estimating, we now assume that we have some prior information from a previous estimation.  Such a prior distribution $P_{\rm prior}(\Omega)$ may be computed either from the RMS error, or from the uncertainty of the previous measurement, plus the typical expected drift uncertainty from the time-varying parameter.  In these cases we can improve MLE by incorporating this information.

The maximum, {\it a posteriori}, probability of the parameter, taking into account the prior is given by Bayes rule, given the measurement record $r(t)$,
\begin{eqnarray}
P(\Omega | r(t) ) &=& \frac{P( r(t) | \Omega) P_{\rm prior}(\Omega)}{P(r(t))}  \nonumber \\
&\propto& P( r(t) | \Omega) P_{\rm prior}(\Omega).
\end{eqnarray}
We may then modify our log-likelihood function as
\be
{\cal L} \propto \ln P( r(t) | \Omega) + \ln P_{\rm prior}(\Omega),
\ee
where the first term is our previous log-likelihood, and the second term takes into account the information from the prior experiments.  
Note that the denominator in Bayes rule can be dropped given that it is independent of $\Omega$, the quantity over which we are maximizing.

\subsection{Time-dependent frequency tracking}

We now illustrate the method of time-dependent parameter tracking of a slowly drifting Rabi frquency.  We first generate a sample drifting frequency on a time scale of longer than 40~$\mu$s, irregularly changing by around 40\% of the value of the Rabi frequency in total.  We then generate the measurement results, sampling from distributions using that slowly changing single qubit Hamiltonian.  This provides a single realization, simulating a typical experiment of this type, where we have included the nonidealities discussed in Sec.~\ref{sec:non-idealities} for realistic values of $T_1 = 50\, \mu$s and $T_2 = 30\, \mu$s, and efficiency of $\eta = 0.5$.  

\begin{figure}[t]
\includegraphics[width=\columnwidth]{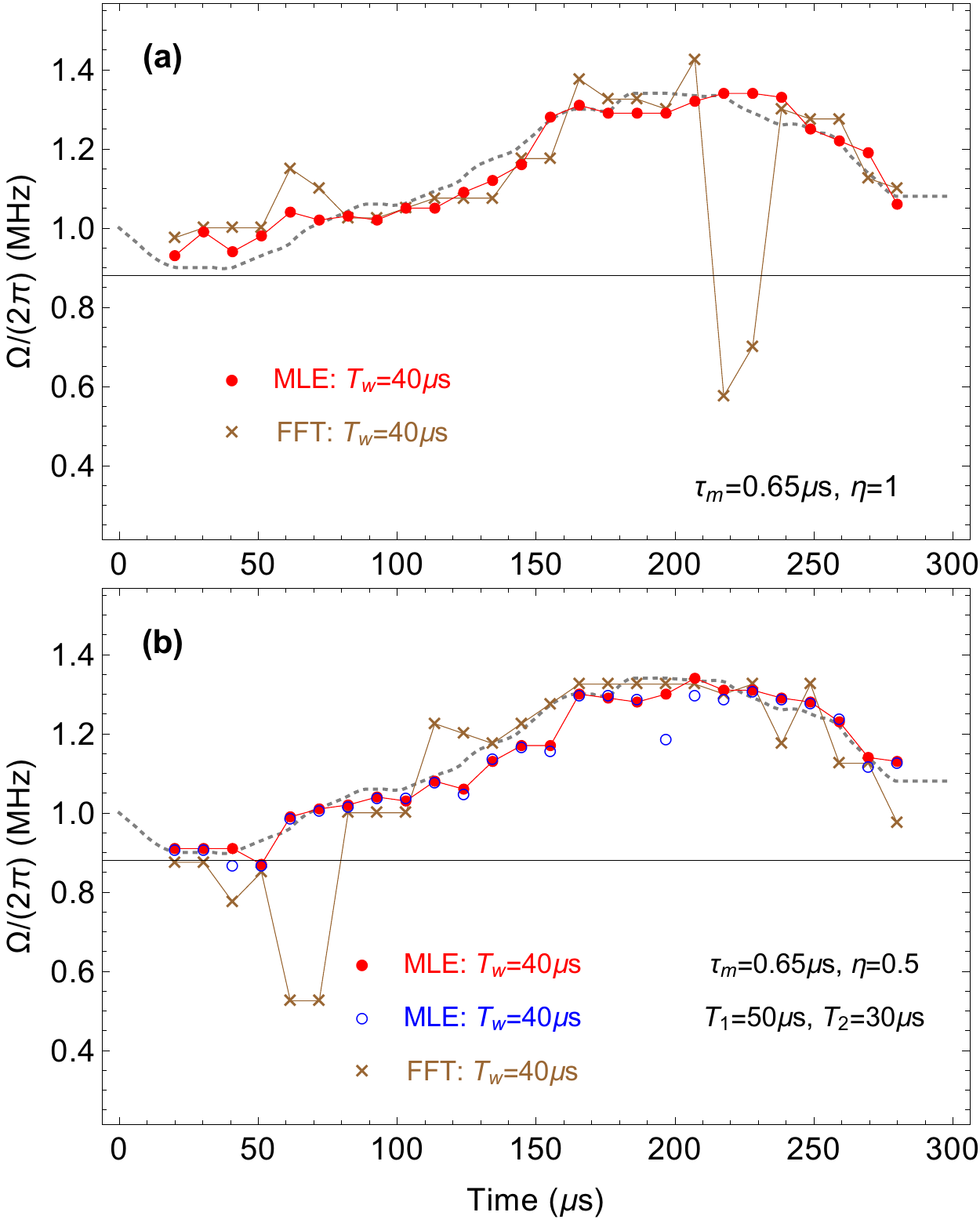}
\caption{The time-dependent frequency tracking, shown in two examples: (a) an ideal case with unit efficiency $\eta = 1$ and no extra dephasing, (b) a realistic case with $50\%$ efficiency ($\eta = 0.5$), a phase relaxation time $T_2 = 30 \mu s$ and an energy relaxation time $T_1 = 50 \mu s$. The true drifting frequency in both plots is shown in dotted gray lines. In the top panel (a), brown crosses represent estimation from FFT and red solid circles represent estimation from MLE. In the bottom panel (b), brown crosses, red solid circles, and blue open circles represent estimation from FFT, MLE, and MLE ignoring nonidealities in the simulation, respectively. The moving window duration is $T_w = 40 \mu s$ for all estimation methods, stepped in $10 \mu s$ intervals. The estimated frequencies are plotted at the mid-point of each estimation window.}
\label{fig-tdft}
\end{figure}

From this data set, we then apply the two discussed estimation strategies, FFT and MLE, with a moving window duration of $T = 40\, \mu$s, matched to the smallest time scale we wish to resolve. We choose the measurement time $\tau_m$ in order to optimize the statistical uncertainty for this time window choice. From Fig.~\ref{contourMLE}, we make the choice $\tau_m = 0.65\mu s$. As shown in Fig.~\ref{fig-tdft}, the methods are able to track this drifting frequency, to the expected accuracies discussed in the previous sections. 

We have run the MLE methods for both the ideal model, and for the non-ideal model, incorporating the $T_1$, $T_2$, and $\eta$ effects discussed in Sec.~\ref{sec:non-idealities}.  The ideal simulation together with the MLE estimation method (red solid circles) and FFT (brown crosses) are shown in Fig.~\ref{fig-tdft}(a), while the simulation incorporating nonidealities mentioned above
is shown in Fig.~\ref{fig-tdft}(b).  In the later subfigure, two different MLE models are used, the first (blue open circles) assume completely ideal dynamics, while the second (red solid circles) incorporates the non-idealities into the estimation model.  Brown crosses are again using the FFT method.  As expected, the MLE methods performs significantly better than the FFT method.
Generally, both MLE methods show good tracking fidelity, indicating there is not much difference in the estimation precision, despite the complexity increase in the non-ideal model.  We explain this insensitivity as a combination of fast Rabi frequency compared to the relaxation rate $1/T_1$, with the rapid purification of the measurement compensating for the dephasing rate $1/T_2$ and inefficiency. The quality of the tracking fidelity varies from run to run, and can always be improved by increasing the duration of the time window.



\section{Conclusion}
We conclude that it is possible to track a drifting parameter with continuous monitoring, and have specified a method that improves the computational overhead to carry out the estimation analysis.  There is a trade off between duration of the data window and of estimation accuracy.  Given a target precision of the estimate, we have shown that this sets the temporal resolution on the drifting parameter, and illustrated that it can work by simulating measurement results.  
If the parameter is changing in an irregular fashion, a major conclusion is that weak continuous monitoring can faithfully track the parameter, and that the method we have described here is statistically optimal for that measurement strategy. We also showed that the method is computationally efficient for such tracking applications, scaling only linearly with the number of time points used in the estimation once an initial frequency range has been identified. This capability opens the door for real-time parameter tracking combined with adaptive feedback \cite{Vijay2012} for parameter stabilization.

The question of how to extend these methods to multiple quantum systems, and estimate parameters such as an interaction energy is an important open question that will be pursued in subsequent work.

{\it Note added:}
After the public presentation of these results at the APS March meeting \cite{March}, but before the posting of our preprint, an independent work covering some of the same physics was posted on the arXiv, authored by Kiilerich and M{\o}lmer \cite{KM4}.

\acknowledgements
We thank Irfan Siddiqi, Alexander Korotkov and Shengshi Pang for helpful discussions.
This work was supported by US Army Research Office Grants No. W911NF-15-1-0496, No. W911NF-13-1-0402, by National Science Foundation grant DMR-
1506081, by John Templeton Foundation grant ID 58558,
by Development and Promotion of Science and Technology Talents Project Thailand, and
by the National Council of Science and Technology (CONACyT) Mexico.
We also acknowledge partial support by Perimeter Institute for Theoretical Physics. Research at Perimeter Institute is supported by the Government of Canada through Industry Canada and by the Province of Ontario through the Ministry of Economic Development \& Innovation.

$^*$  L.C. and A.C. contributed equally to this work.

\section{Appendix: Maximum Likelihood Method for Periodic Projective Measurements}
In this appendix we review how to find the maximum likelihood estimate for the qubit oscillation frequency, using only unitary evolution and periodic projective measurements.  Starting in basis state $|0\ra$, the unitary (2) followed by projective measurement will yield either result 0, or 1, so the system has probability $p_s = \cos(\Omega \tau/2)^2$ to be projected back into state $|0\ra$ (given result 0), and probability $p_d = \sin(\Omega \tau/2)^2$ to be projected into state $|1\ra$ (given result 1).  Similarly, starting in the basis state $|1\ra$, following the unitary operation and measurement, the system has probability $p_s = \cos(\Omega \tau/2)^2$ to be projected back into state $|1\ra$ (given result 1), and probability $p_d = \sin(\Omega \tau/2)^2$ to be projected into state $|0\ra$ (given result 0).  Thus the system has probability $p_s$ of staying the same as the previous step, and probability $p_d$ of being different from the previous step after the measurement.

We now are given a sequence of 0s and 1s resulting from a lengthy number of measurements. Suppose we also know the initial state of the system.  How can we extract the value of $\Omega$ if we do not know it in advance? We see from the discussion above that the probability of a sequence of results can be determined by the number of switches from the same result to a different result, $n$.  Given $N$ measurements, the total probability is
\be
P(n, N|\Omega) = \binom{N}{n} (\sin^2 \Omega \tau/2)^n (\cos^2 \Omega \tau/2)^{N-n},
\ee
where the prefactor normalizes the distribution.  This is just a binomial probability distribution with probability $p_d$ of switching and probability $p_s = 1-p_d$ of staying the same.  In this simple example the Fisher information about the parameter $\Omega$ may be calculated straightforwardly,
\be
{\cal I} = \sum_{n=0}^N P(n, N|\Omega) \left[\partial_\Omega \ln P(n, N|\Omega)\right]^2 = N \tau^2,
\ee
so the standard deviation is bounded by the Cram\'er-Rao bound, $\sigma_\Omega \ge 1/(\tau \sqrt{N})$. 

We may now use the maximum likelihood method (MLE) to find an estimate of $\Omega$ and the uncertainty in the estimate.  We expand the log-likelihood as a function of $\Omega$ as
\begin{eqnarray}\label{eq-expandlogp}
\ln P &\sim&  const + \frac{\partial}{\partial \Omega} \ln P(n, N|\Omega_{ML}) (\Omega - \Omega_{ML}) \\
&+& (1/2)\frac{\partial^2}{\partial \Omega^2} \ln P(n, N|\Omega_{ML}) (\Omega - \Omega_{ML})^2 + \cdots, \nonumber
\end{eqnarray}
and define $\Omega_{ML}$ as the frequency that maximizes it. In turn, the second derivative evaluated at $\Omega = \Omega_{ML}$ gives the (negative) inverse uncertainty variance in the estimate. 
This yields
\be
\tau\, \Omega_{ML} = 2\arcsin \sqrt{\frac{n}{N}},
\ee
with a standard deviation uncertainty of $\sigma_{\Omega} =  1/(\tau \sqrt{N})$, indicating that the maximum likelihood method saturates the Cram\'er-Rao bound.   We note that although the uncertainty is independent of $n$, the estimate has a divergent slope at $n=0$ and $n=N$, with a minimum slope at $n=N/2$, indicating that the estimate is least sensitive to fluctuations in $n$ at $\tau \Omega = \pi/2$.
We have checked this method by simulating the sequence of outputs from such a measurement by first fixing a known value $\Omega$, running the simulation and estimating the parameter from the simulation data. We then compare the estimate to the known value.   This method works robustly for different values of $n$ and $N$, and gives estimated values consistent with the expected uncertainty.

\end{document}